 \documentclass[aps,prl,reprint,superscriptaddress]{revtex4-2}

\makeatletter
\def\maketitle{
\@author@finish
\title@column\titleblock@produce
\suppressfloats[t]}
\makeatother
 
 \usepackage{ragged2e}
\usepackage{amsmath}
\usepackage{amssymb}
\usepackage{physics}
\usepackage{graphicx}%
 \usepackage{setspace}
 \usepackage[skip=5pt]{caption}
\usepackage{dcolumn}
\usepackage{bm}
\usepackage[dvipsnames]{xcolor}

\begin{document}


\title{Novel Electronic Structure of Nitrogen-Doped Lutetium Hydrides}


\author{Adam Denchfield}
\affiliation{Department of Physics, University of Illinois Chicago, Chicago, Illinois 60607, USA}


\author{Hyowon Park}
\affiliation{Department of Physics, University of Illinois Chicago, Chicago, Illinois 60607, USA}
\affiliation{Materials Science Division, Argonne National Laboratory, Lemont, Illinois 60439, USA}
\author{Russell J. Hemley}
\affiliation{Department of Physics, University of Illinois Chicago, Chicago, Illinois 60607, USA}
\affiliation{Department of Chemistry, University of Illinois Chicago, Chicago, Illinois 60607, USA}
\affiliation{Department of Earth and Environmental Sciences, University of Illinois Chicago, Chicago, Illinois 60607, USA} 

\date{June 7th, 2023}

\begin{abstract}
   First-principles density functional theory (DFT) calculations of Lu-H-N compounds reveal low-energy configurations of Fm$\overline{3}$m Lu$_{8}$H$_{23-x}$N structures that exhibit novel electronic properties such as flat bands, sharply peaked densities of states (van Hove singularities, vHs), and intersecting Dirac cones near the Fermi energy (E$_F$). These N-doped LuH$_3$-based structures also exhibit an interconnected metallic hydrogen network, which is a common feature of high-T$_c$ hydride superconductors. Electronic property systematics give estimates of T$_c$ for optimally ordered structures that are well above the critical temperatures predicted for structures considered previously. The vHs and flat bands near E$_F$ are enhanced in DFT+U calculations, implying strong correlation physics should also be considered for first-principles studies of these materials. These results provide a basis for understanding the electronic properties observed for nitrogen-doped lutetium hydride.
\end{abstract}


\maketitle


Rare-earth hydrides form a unique chemical composition of the heavy rare-earth ions and the lightest hydrogen atoms that can lead to novel material properties including metal-insulator transitions and superconductivity.
Such novel properties originate from the subtle balance between the strongly localized nature of heavy electrons in rare-earth ions and the itinerant electronic behaviors of hydrogen.
A particularly intriguing example is nitrogen-doped lutetium hydride, which has recently been reported to exhibit near-ambient superconductivity \cite{dasenbrock2023evidence}. The disposition and stoichiometry of the light elements are not known, nor is it clear what role, if any, nitrogen plays in the reported superconductivity. First-principles searches have not found structures with strong electron-phonon coupling (EPC) \cite{huo2023first, hilleke2023structure, ferreira2023search, lu2023electron, sun2023effect, lucrezi2023temperature}. It is of interest to find structures in the ternary Lu-H-N system with strong vHs near the Fermi energy (E$_F$) in the density of states (DOS) because those features increase EPC \cite{sano2016effect, gai2022van}. 

Inspired by the pioneering theoretical predictions of high T$_c$ superconductivity by Ashcroft \cite{ashcroft1968metallic, ashcroft2004hydrogen}, density functional theory (DFT) calculations have played an important role in guiding experiments aimed at realizing room temperature superconductivity, specifically in dense hydrides \cite{liu2017potential, peng2017hydrogen} with calculations that have been confirmed by subsequent experiments \cite{somayazulu2019evidence, drozdov2019superconductivity, snider2021synthesis, kong2021superconductivity, troyan2021anomalous} (see \cite{hilleke2022tuning} for a review). The recent report of room-temperature superconductivity in LuH$_{3-x}$N$_y$ at 1 GPa \cite{dasenbrock2023evidence} has motivated numerous recent DFT calculations aimed at understanding the result \cite{dasenbrock2023evidence, huo2023first, hilleke2023structure, ferreira2023search, xie2023lu, lucrezi2023temperature, lu2023electron}. Candidate structures and stoichiometries have been identified and proposed, but they do not exhibit properties conducive to known superconducting mechanisms. We use DFT \cite{kohn1965self, giannozzi2009quantum, giannozzi2017advanced, giannozzi2009quantum} and DFT+U \cite{wang2016local, tolba2018dft+} calculations to explore bonding and electronic properties of a broad range of structures and compositions that fit the available experimental data. We show that a subset of energetically preferred Lu-H-N structures and stoichiometries consistent with available experimental constraints \cite{dasenbrock2023evidence} exhibit novel electronic structures that include Dirac cones, a sharp vHs and flat bands at E$_F$. These features may underlie the remarkable near-ambient superconductivity reported in the Lu-H-N system \cite{dasenbrock2023evidence}. 

\begin{figure}
   \centering
     \includegraphics[width=0.9\linewidth]{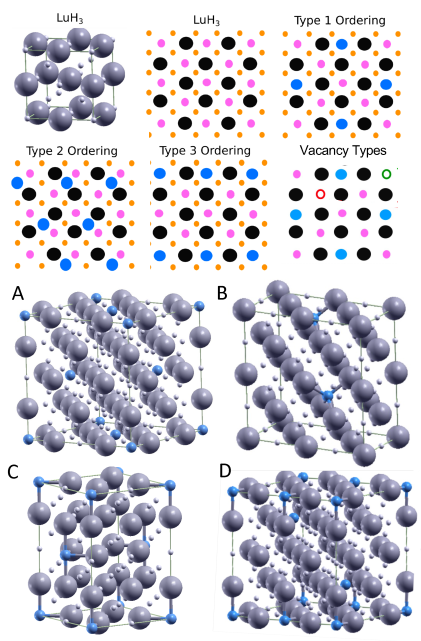}
   \caption{\justifying Top: Unit cell of Fm$\overline{3}$m LuH$_3$ followed by depictions of the layers [(100) planes] that form the building blocks of the superlattice structures considered in the present study: LuH$_3$ layer; Type 1 ordering, with N in the octahedral site; Type 2 ordering, with N in the tetrahedral site, Type 3 ordering, with chains of octahedrally coordinated N; and the two types of octahedral H vacancies denoted by open circles (red: type 1, purple: type 2), with the tetrahedral hydrogens omitted for clarity. Bottom: the four 2 x 2 x 2 superlattice structure-types for Lu$_8$H$_{23-x}$N (A, B, C, D) created from the above ordering schemes (see text and SI).}
   \label{fig:NFCC_struct_1}
 \end{figure}

We begin our study by computing the electronic structure of a variety of supercells based on Fm$\overline{3}$m LuH$_{3-x}$N$_{y}$ in light of recent theoretical studies of the energetics of the system \cite{huo2023first, hilleke2023structure, ferreira2023search, xie2023lu, lucrezi2023temperature, lu2023electron}. In the parent structure Fm$\overline{3}$m LuH$_3$, each Lu atom is surrounded by hydrogens in tetrahedral and octahedral sites (Fig.\ \ref{fig:NFCC_struct_1}, top). Substituting N atoms in hydrogen locations, we find that N atoms prefer being spatially separated (see Table S1) due to competition over the Lu$_d$ electrons with which they bond. This also minimizes the Coulomb interaction between N atoms. This spatial separation as well as the experimentally constrained 1:8 N:Lu ratio \cite{dasenbrock2023evidence} (i.e., $y=0.125$ in LuH$_{3-x}$N$_y$) yields a subset of energetically favorable structures containing highly symmetric configurations. Fm$\overline{3}m$ Lu$_{8}$H$_{23-x}$N for small $x$ emerged as particularly important due to their interesting electronic structure and for consistency with experiments upon relaxation (Fig.\ S6). 

We consider four main superlattice structure types with stoichiometry Lu$_{8}$H$_{23-x}$N (Fig.\ \ref{fig:NFCC_struct_1}, bottom). Structure A consists of alternating layers of LuH$_3$ and type 1 ordering LuH$_{2.75}$N$_{0.25}$, resulting in a face-centered-cubic (FCC) superlattice of N atoms. Structure B has N in tetrahedral sites (type 2 ordering), while A,C, and D have them in octahedral sites. Structure C has an ABCB stacking, where B are LuH$_3$ layers, and A and C layers are type 1 orderings with a relative shift. Structure D is more complex, but its key feature is N-Lu-N chains. The full structural information is presented in Figs. S1-S5. To study hydrogen vacancies, we present two inequivalent octahedral hydrogen vacancy positions inside the type 1 ordering scheme. The unit cell parameters and atomic positions of the superlattice structure-types were relaxed without symmetry (see SI). When the above structures are relaxed, the Lu atoms move towards the N atoms. The simulated x-ray diffraction patterns of the structures are broadly consistent with those measured experimentally \cite{dasenbrock2023evidence} (see Fig.\ S6). However, important superlattice reflections at low angle were not observed because of limitations of the experiment. The optimized lattice parameter of structure A (Fig.\ S1) at zero pressure with DFT-PBE ($a = $ 10.03 \AA) is close to twice the value obtained experimentally for the dominant Fm$\overline{3}$m phase in the experiments ($a = $ 5.023 \AA) \cite{dasenbrock2023evidence}.

\begin{figure}
  \centering
  \includegraphics[width=1.0\linewidth]{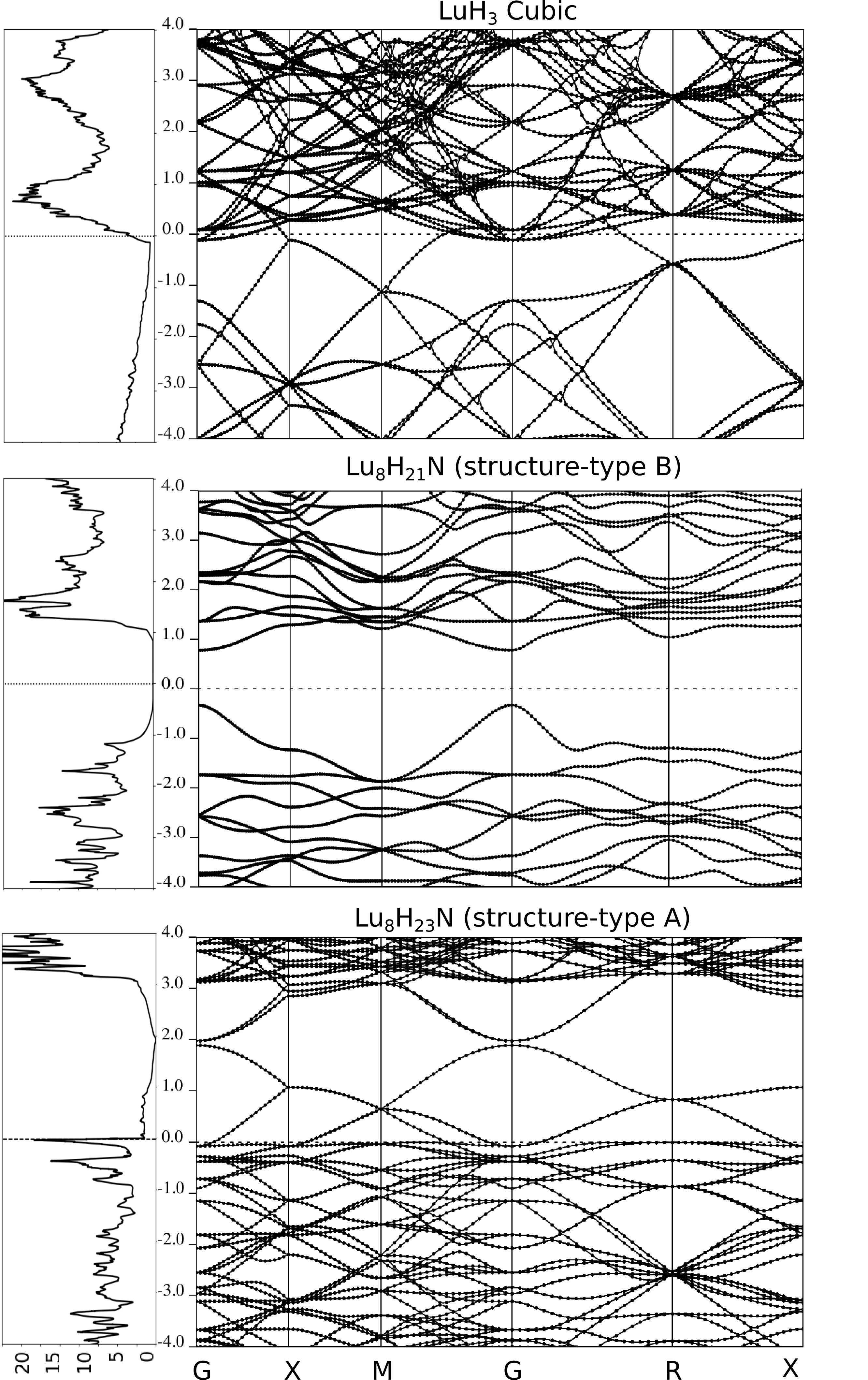}
  
  \caption{\justifying The band structures of (top) Fm$\overline{3}$m LuH$_{3}$, (middle) Lu$_8$H$_{21}$N (Lu$_8$H$_{21}$N structure-type B) \cite{sun2023effect}, and (bottom) Lu$_8$H$_{23}$N (structure-type A) using DFT+U, U=8.2 eV on Lu$_d$, 5.5 eV on Lu$_f$.}
  \label{fig:compare_bands}
\end{figure}

Figure \ref{fig:compare_bands} shows the band structures of Fm$\overline{3}$m LuH$_3$, Lu$_8$H$_{21}$N (structure-type B) \cite{sun2023effect}, and Lu$_8$H$_{23}$N (structure-type A), the latter using DFT+U (see also Fig.\ S7). Our computed LuH$_3$ band structure is in good agreement with the DFT results calculations of Sufyan and Larsson \cite{sufyan2023topological} (Fig.\ S8). Lu$_8$H$_{21}$N (B) has a sizeable band gap, meaning it cannot be a superconducting phase. Despite the same number of valence electrons, this large difference with LuH$_3$ is likely due to the stoichiometry rather than differences in tetrahedral vs octahedral site chemistry. Placing N in tetrahedral vs octahedral positions does not significantly alter the DOS, seen by comparing all the PDOS in Fig.\ S9(a-d). Sun et al \cite{sun2023effect} examined the stability and properties of Lu$_8$H$_{21}$N (B) and found it is not conducive to superconductivity \cite{sun2023effect}, consistent with our calculated band structure. Remarkably, the bands of Lu$_8$H$_{23}$N (structure A) are distinctly different, exhibiting a combination of hybridization and correlation effects giving rise to a very sharp vHs at E$_F$ and regions of nearly flat bands (discussed later) with sharp intersecting Dirac cones (see Fig.\ S10). 

\begin{figure}
    \centering
    \includegraphics[width=1.0\linewidth]{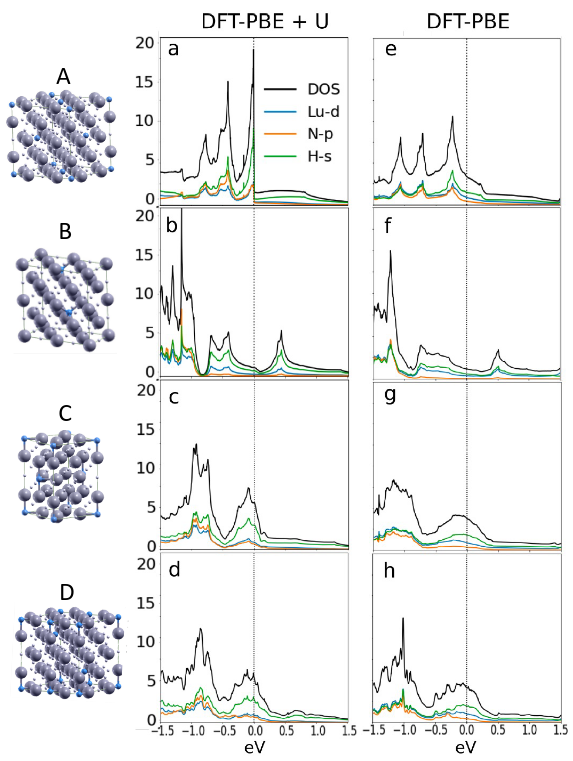}
    \caption{\justifying Calculated PDOS near E$_F$ of superlattice structure types A-D with DFT-PBE+U, U=8.2 eV on Lu$_d$ \cite{topsakal2014accurate} (a-d) vs DFT-PBE (e-h). }
    \label{fig:PDOS_compare_many}
\end{figure}

Figure \ref{fig:PDOS_compare_many} compares the partial density of states (PDOS) near E$_F$ of these and other structures including both DFT+U and DFT results. 
We placed U=$8.2$ eV on the Lu$_d$ orbitals since the pseudopotentials used are tested for this U value to match accurate (but expensive) HSE06 functional calculations on LuN \cite{topsakal2014accurate}. Lu$_8$H$_{23}$N structure A exhibits an extraordinarily sharp vHs close to E$_F$ [Fig.\ \ref{fig:PDOS_compare_many}(a); see also Fig.\ S9] that is dominated by H$_s$ states with additional contributions from N$_p$ and Lu$_d$. It is known that sharp vHs at E$_F$ lead to T$_c$ estimates that grow unbounded with the DOS at E$_F$ \cite{cappelluti1996nonadiabatic, sano2016effect, bok2012superconductivity} unlike the flat DOS approximation \cite{bardeen1957theory}. 
 Lu$_8$H$_{23}$N structures B-D are relatively featureless around E$_F$ [Fig.\ \ref{fig:PDOS_compare_many}(b-d)] but share similar broad features below E$_F$ and become more similar when unrelaxed (Fig.\ S9). The crystal field splitting and H$^{\text{oct}}_s$, H$^{\text{tet}}_s$ projections for the parent structure LuH$_3$ and structures A-D are studied in the SI [Figs.\ S7(top), S9]. 
 
 To study the correlation effects we compare the DFT-PBE results in Fig.\ \ref{fig:PDOS_compare_many}(e-h). The removal of U changes structure A, whose vHs is now broadened and 0.2 eV below E$_F$ [Fig.\ \ref{fig:PDOS_compare_many}(e)]. The removal of U for the other structures changes their DOS marginally [Fig.\ \ref{fig:PDOS_compare_many}(f-h)]. The main effect of U in Lu$_8$H$_{23}$N (A) is to push Lu$_d$ states away from E$_F$ [see Fig.\ S11(left)], but near E$_F$ multiband effects complicate matters [Fig.\ S11(right)] resulting in an enhancement of the vHs at E$_F$ [Figs.\ \ref{fig:PDOS_compare_many}(a) and S12]. Ideally, the U value is computed for each different structure, which complicates large structure searches. This may explain why structure type A compounds have been missed so far in first-principles calculations only using DFT-PBE.
 
In all the configurations shown the octahedral hydrogens are the primary contributors, with additional Lu$_d$, tetrahedral H, and N$_p$ contributions (Fig.\ S9). The results contrast with those of Ferreira et al \cite{ferreira2023search} who found only structures with primarily Lu$_d$ character at E$_F$. Of the structures considered thus far, structure A uniquely obtains a sharp vHs at E$_F$ with DFT+U. Exploring the effect of introducing vacancies into structure A on the vHs (Fig.\ S13), we find the electronic structure (vHs) is stable to a small number of octahedral hydrogen vacancies with the effect of raising E$_F$. This in combination with the hole-doping effect of N atoms on the LuH$_3$ bands in Lu$_8$H$_{23}$N (see Fig.\ S7) can thus tune E$_F$ to match the van Hove singularity peak, an effect that can significantly enhance superconductivity in other materials \cite{ge2020hole, gai2022van}. In addition, the application of modest pressure (i.e. 2 GPa) to Lu$_8$H$_{23}$N (A) increases the vHs energy by 13 meV, potentially changing T$_c$ significantly over this range as the vHs goes from below E$_F$ to above it. These options represent a strategy to tune the parameters of Lu$_8$H$_{23-x}$N (A) to maximize the achievable T$_c$. 

The total energies of the Lu$_8$H$_{23}$N structures considered here (Table S1) are within 7-30 meV/atom of each other. The DFT results are consistent with those reported for Lu$_8$H$_{23}$N by Sun et al \cite{sun2023effect} who also found Lu$_8$H$_{21}$N (B) to be dynamically stable. Calculations for selected structures considered here suggest any dynamical instabilities are weak. On the other hand, these instabilities are calculated in the harmonic and Born-Oppenheimer approximations, and are likely outweighed by nuclear quantum effects and anharmonicity as found for other hydrides \cite{errea2015high, liu2018dynamics, errea2020quantum, ly2022stability}. We also considered the effects of magnetic properties on the energetics. DFT and DFT+U calculations indicate any magnetic states have small magnetic moments with energies lowered by less than 1 meV/atom, i.e., insignificant compared to ambient temperatures. 

\begin{figure}
    \centering
    \includegraphics[width=1.0\linewidth]{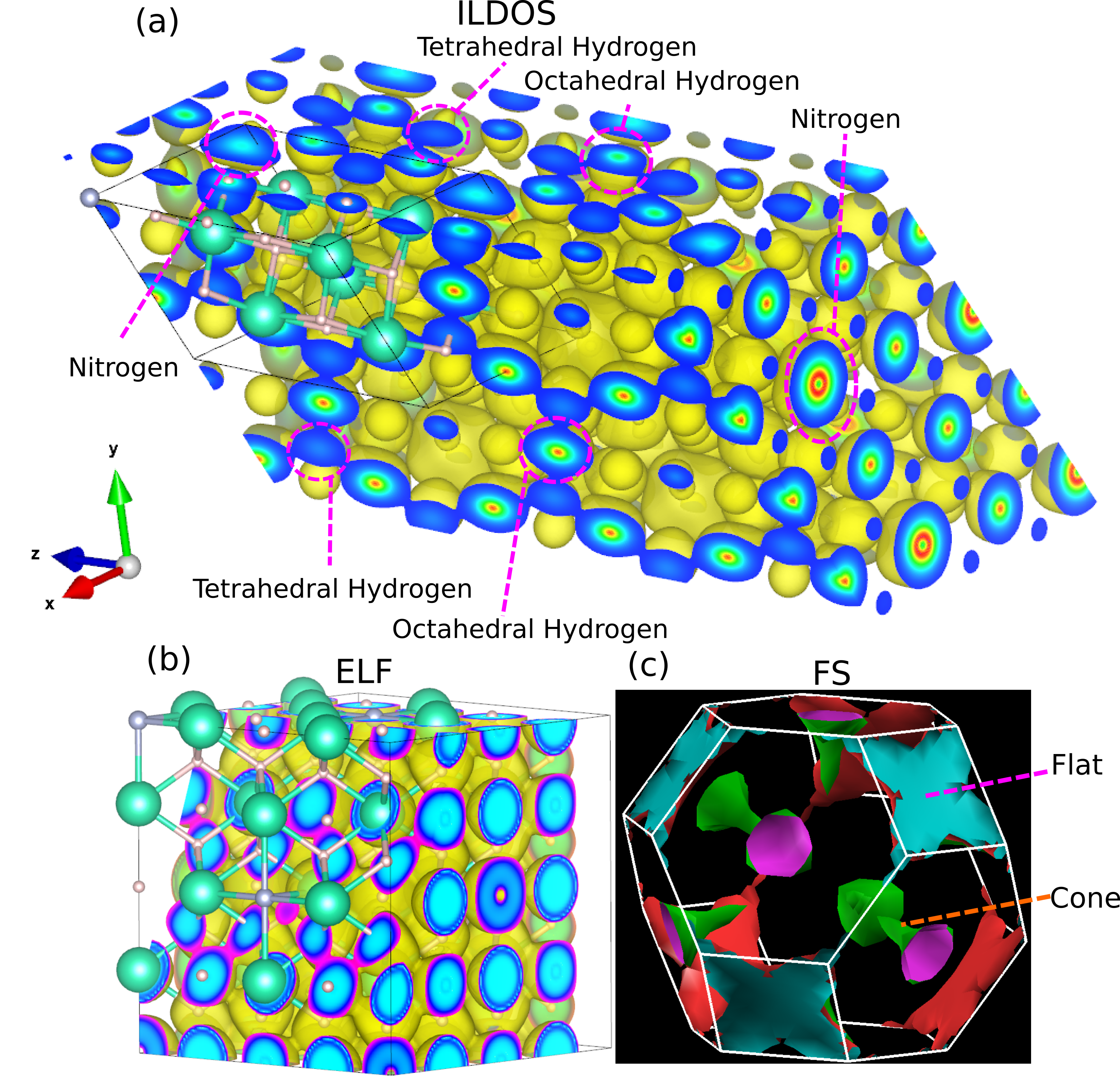}
    \caption{\justifying Calculated local bonding and electronic properties for Lu$_8$H$_{23}$N (A), with DFT+U, U on Lu$_d$ of 8.2 eV. (a) Integrated local density of states (ILDOS), integrated $\pm 5$ meV from E$_F$. (b) The ELF, illustrating the formation of the hydrogen bonding network. (c) Fermi surface (FS), showing both flat and conical bands. }
    \label{fig:FS_ILDOS_ELF}
\end{figure}

Figure \ref{fig:FS_ILDOS_ELF} further illustrates the electronic properties near E$_F$ of Lu$_8$H$_{23}$N (A) with DFT-PBE+U. The integrated local density of states (ILDOS), integrated around the Fermi energy, shows that the conduction states are composed of octahedral hydrogen and N states, with nontrivial contributions from the tetrahedral hydrogen. A ``metallic hydrogen" network forms in the ILDOS (sliced three ways for illustrative purposes) with mostly-localized N states; smaller and larger rings of octahedral and tetrahedral hydrogen are also observed. Given the ``metallic hydrogen" network, we also plot the electron localization function (ELF)\cite{savin1997elf, fuentealba2007understanding} and find that the hydrogen ELF spheres begin to overlap at an isovalue of 0.51. Belli et al \cite{belli2021strong} found that the ELF and relative H PDOS at E$_F$ correlates with the values of theoretically predicted T$_c$ of hydride superconductors. Use of their correlation gives a T$_c$ in the 100 K range for this structure. A full Eliashberg calculation is beyond the scope of the present study due to the computational resources needed for the supercells considered here. Nevertheless, the EPC enhancement from vHs and flat bands at E$_F$ found for other high-T$_c$ superconductors \cite{sano2016effect, gai2022van, liu2017potential} suggests that the band structure characteristics of Lu$_8$H$_{23-x}$N phases predicted here support the existence of strong EPC (and thus very high T$_c$) for the Lu-H-N system studied in Ref.\ \cite{dasenbrock2023evidence}. 

We also comment on the topological character of band structures found for these LuH$_{3-x}$N$_y$ phases. We confirm that Fm$\overline{3}$m LuH$_3$ has vHs and Dirac cones near E$_F$ and is thus a potential Dirac semimetal, in agreement with recent calculations of Sufyan and Larsson \cite{sufyan2023topological}. The incorporation of N to form Lu$_8$H$_{23}$N (A) shifts those topological features of LuH$_3$ 2 eV above E$_F$ (see Fig.\ S7). The Fermi surface (FS) of Lu$_8$H$_{23}$N (A) (Fig. \ref{fig:FS_ILDOS_ELF}) clearly shows the flat band regions and the Dirac cones. The flat bands of Lu$_8$H$_{23}$N (A) are intersected by other bands (Figs.\ \ref{fig:compare_bands} and S10), which can result in non-trivial topology for the bands (such as a nonzero Chern number). This is a necessary ingredient for flat-band superconductivity \cite{peotta2015superfluidity} (also  considered topological superconductivity \cite{sato2017topological}). The band structure also has interesting parallels with the flat bands predicted for certain dense hydrogen structures of \cite{naumov2013graphene}. Nearly flat bands can emerge in multiband tight-binding models \cite{sun2011nearly, leykam2018artificial} due to geometric hybridization effects on idealized Kagome, Lieb, and checkerboard lattices \cite{kang2020topological}. These structures all share the trait of being highly symmetric, like those considered here. The nearly flat bands of Lu$_8$H$_{23}$N (A) (Figs. \ref{fig:compare_bands} and S10) likely arise from the relative phases of the hopping parameter, and are further flattened with DFT+U. 

Finally, our results have important implications for experimental studies of the reported near-ambient superconductivity in the Lu-H-N system \cite{dasenbrock2023evidence}. The sensitivity of the electronic properties in Lu$_8$H$_{23-x}$N based phases to nitrogen and hydrogen-vacancy ordering suggests that the reported high-T$_c$ superconductivity as well as other emergent phenomena will be highly dependent on sample preparation and annealing. This finding may explain the apparent difficulties reported \cite{ming2023absence, cai2023no, xing2023observation} in reproducing the results presented in Ref. \cite{dasenbrock2023evidence}. In addition, our calculated diffraction patterns should help guide experimental studies of the crystal structure (Fig. S6). We also note the that the temperature dependence of the electrical resistance reported for Lu-H-N samples \cite{dasenbrock2023evidence} shows striking parallels to anomalous resistivity curves documented many years ago for other substoichiometric lanthanide trihydrides (LnH$_{3-x}$) across their metal-insulator transitions \cite{shinar1990anomalous, libowitz1972electronic} (Fig.\ S14), where the electrical conductivity has been shown to be strongly dependent on hydrogen-vacancy ordering \cite{kerscher2012first}.

In summary, we have identified classes of structures, stoichiometries, and hydrogen-vacancy ordering schemes for Lu-H-N with remarkable electronic properties. We predict that strong correlations give rise to structures with flat band physics at E$_F$, which are expected to have high EPC \cite{sano2016effect, gai2022van}. We also predict that the proposed structures will exhibit interesting correlated physics in addition to their potential for superconductivity, due to the flat bands and intersecting Dirac cones. Confirmation of the predicted structures awaits detailed single-crystal x-ray and neutron scattering experiments, while the interesting band physics can be examined using various electron and x-ray spectroscopies possible at ambient and near-ambient conditions. 

 We thank E. Zurek and S. A. Gramsch for useful discussions. 
 This research was supported by the NSF (DMR2104881) and DOE-NNSA through the Chicago/DOE Alliance Center (DE-NA0003975) (AD and RH), and the Materials Sciences and Engineering Division, Basic Energy Sciences, Office of Science, US Department of Energy (HP). 
We gratefully acknowledge the computing resources provided on Bebop, a high-performance computing cluster operated by the Laboratory Computing Resource Center at Argonne National Laboratory.

\bibliography{combined}
\vspace{2cm} 
\newpage
\clearpage

\section*{SUPPLEMENTAL INFORMATION}

\renewcommand{\thefigure}{S\arabic{figure}}
\setcounter{figure}{0}
\subsection*{Methods}

\label{sec:methods}
We performed DFT calculations primarily with the \texttt{Quantum Espresso} software package \cite{giannozzi2017advanced}. The SSSP database \cite{prandini2018precision} was used to vet pseudopotentials for convergence of the computed pressure, phonon frequencies, and formation energies. Accordingly, we chose the pseudopotential for Lu from Ref. \cite{topsakal2014accurate}, and the \texttt{pslibrary} pseudopotentials for N and H \cite{dal2014pseudopotentials}. We built supercells of LuH$_{3-x}$N$_y$ with structures up to 128 atoms and relaxed them using DFT-PBE. We performed variable-cell calculations for the relaxation without symmetry constraints, and perturbed all atoms slightly for the relaxation to break any symmetries that would prevent relaxation to a local minimum (see Figs.\ \ref{fig:typeA_struct}-\ref{fig:typeD_struct}). Each relaxed structure had remaining forces under \texttt{1e-5} Ry/Bohr, and energy differences were under \texttt{1e-5}Ry. The x-ray patterns were simulated using pymatgen \cite{ong2013python} and use Cu-K$\alpha$ radiation for comparison with recent experiments \cite{dasenbrock2023evidence}. We used wavefunction plane-wave cutoffs of 1000 eV but performed convergence tests up to 1154 eV on many of the structures and verified they yielded the same results. We generally used tight k-meshes corresponding to 0.13 \AA$^{-1}$, but did convergence tests with more k-points as needed, including on the vHs of Lu$_8$H$_{23}$N (see Fig.\ \ref{fig:kmesh_vHs}). We also used the optimized tetrahedron method \cite{kawamura2014improved}. The \texttt{VESTA} code \cite{momma2011vesta} was used for the visualizations of the ELF and ILDOS. ILDOS calculations performed integrations within a range of 5meV around E$_F$. We used \texttt{Quantum Espresso}'s default atomic projections for the PDOS calculations. Using the projector-augmented-wave (PAW) projectors reduces the hydrogen weight and implies N$_p$ is the dominant term for the PBE calculations. However, the PAW projectors are zero outside of the PAW sphere and thus may underestimate contributions from dispersive orbitals like H$_s$. We also performed comparisons of the PBE functional with the vdW-DF3-opt2 functional \cite{chakraborty2020next} for many of the calculations and saw no meaningful changes. We performed DFT+U calculations with the U-values for Lu$_d$ recommended in Ref. \cite{dal2014pseudopotentials}, which vetted the pseudopotentials and U-values with calculations on LuN using the same codes and functional we use. The ideal U-value is not tied to the atom, and in principle it should be structure-dependent. Nevertheless, we use a fixed U-value for Lu$_d$ (except for the calculations shown in Fig. \ref{fig:four_cases}) as it is generally difficult to find the ideal U-value for each structure. Ref. \cite{topsakal2014accurate} also recommends U=5.5 eV for Lu$_f$ electrons but use of this value did not significantly change the electronic properties near E$_F$. 

\setcounter{figure}{0}
\begin{figure*}[b]
\captionsetup{name={TABLE}}
    \centering
    \caption{\justifying Calculated total energies (eV/atom) of Lu$_8$H$_{23}$N type structures. The unrelaxed structures are all supercells of Fm$\overline{3}$m LuH$_3$ with the appropriate N substitutions; all have lattice constant 9.89 \AA  (see Figs.\ \ref{fig:typeA_struct}, \ref{fig:typeB2_struct}, \ref{fig:typeC_struct}, \ref{fig:typeD_struct} for the relaxed structure parameters).}
    \includegraphics[width=0.7\linewidth]{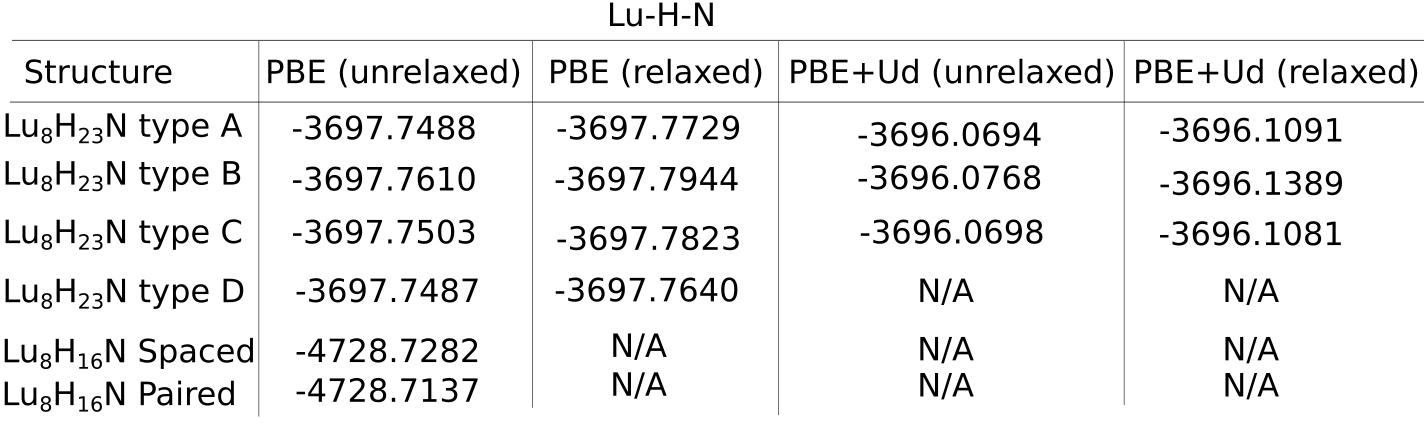}
    \label{fig:energetics}
\end{figure*}
\setcounter{figure}{0}

\begin{figure*}
    \centering
    \includegraphics[width=0.8\linewidth]{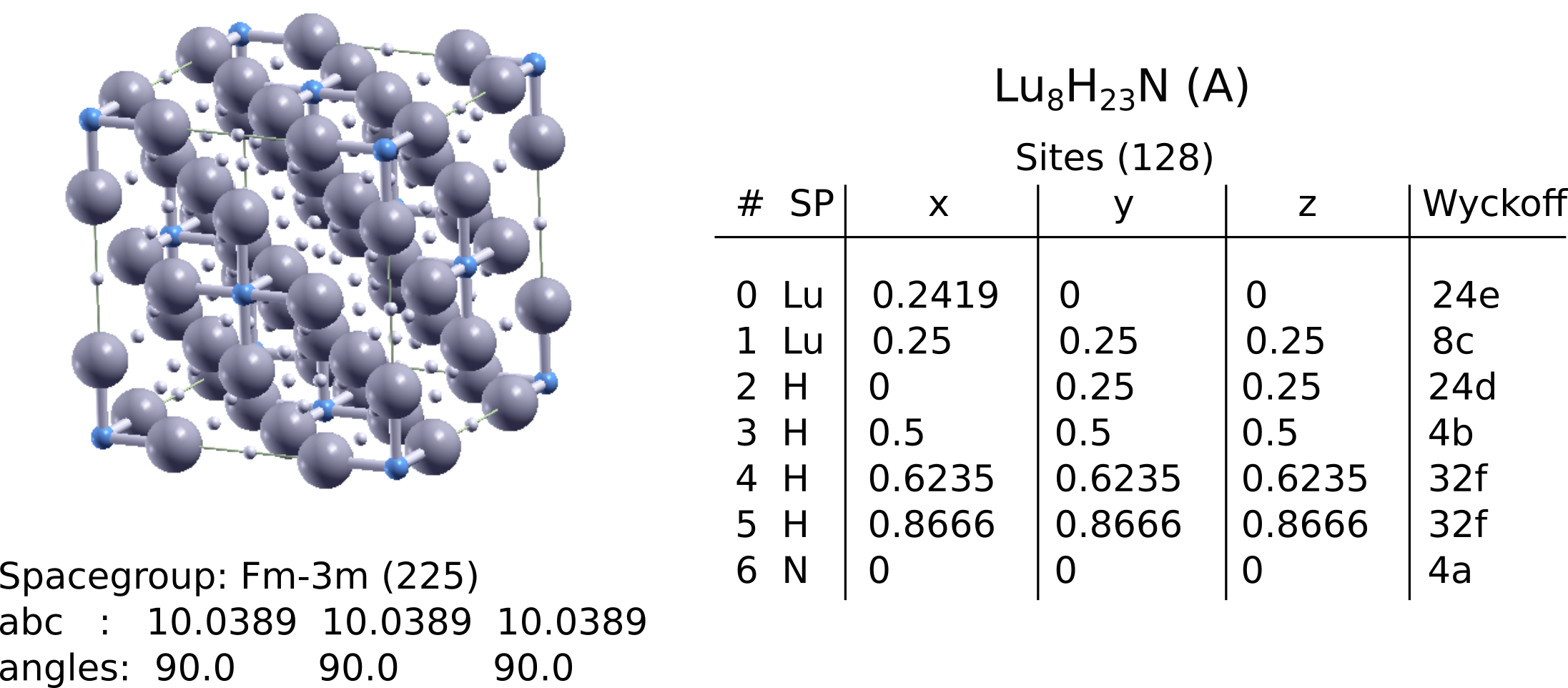}
    \caption{Structure A (Lu$_8$H$_{23}$N), conventional unit cell parameters, and atomic positions.}
    \label{fig:typeA_struct}
\end{figure*}

\begin{figure*}
    \centering
    \includegraphics[width=0.8\linewidth]{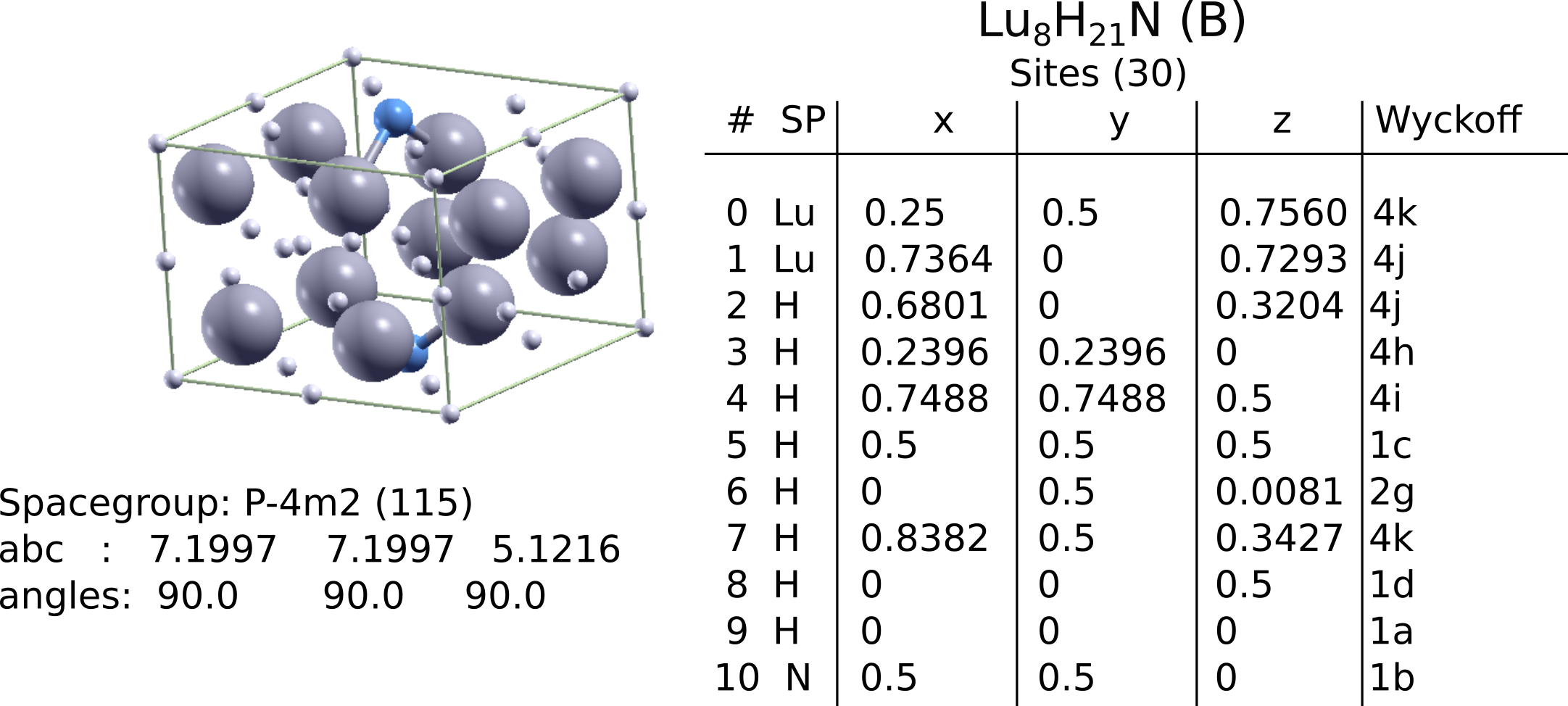}
    \centering \caption{\justifying  Structure B (Lu$_8$H$_{21}$N), conventional cell parameters, and atomic positions. Our relaxed 
    structure is close to that reported in Ref. \cite{sun2023effect}.}
    \label{fig:typeB_struct} 
\end{figure*}

\begin{figure*}
    \centering
    \includegraphics[width=0.8\linewidth]{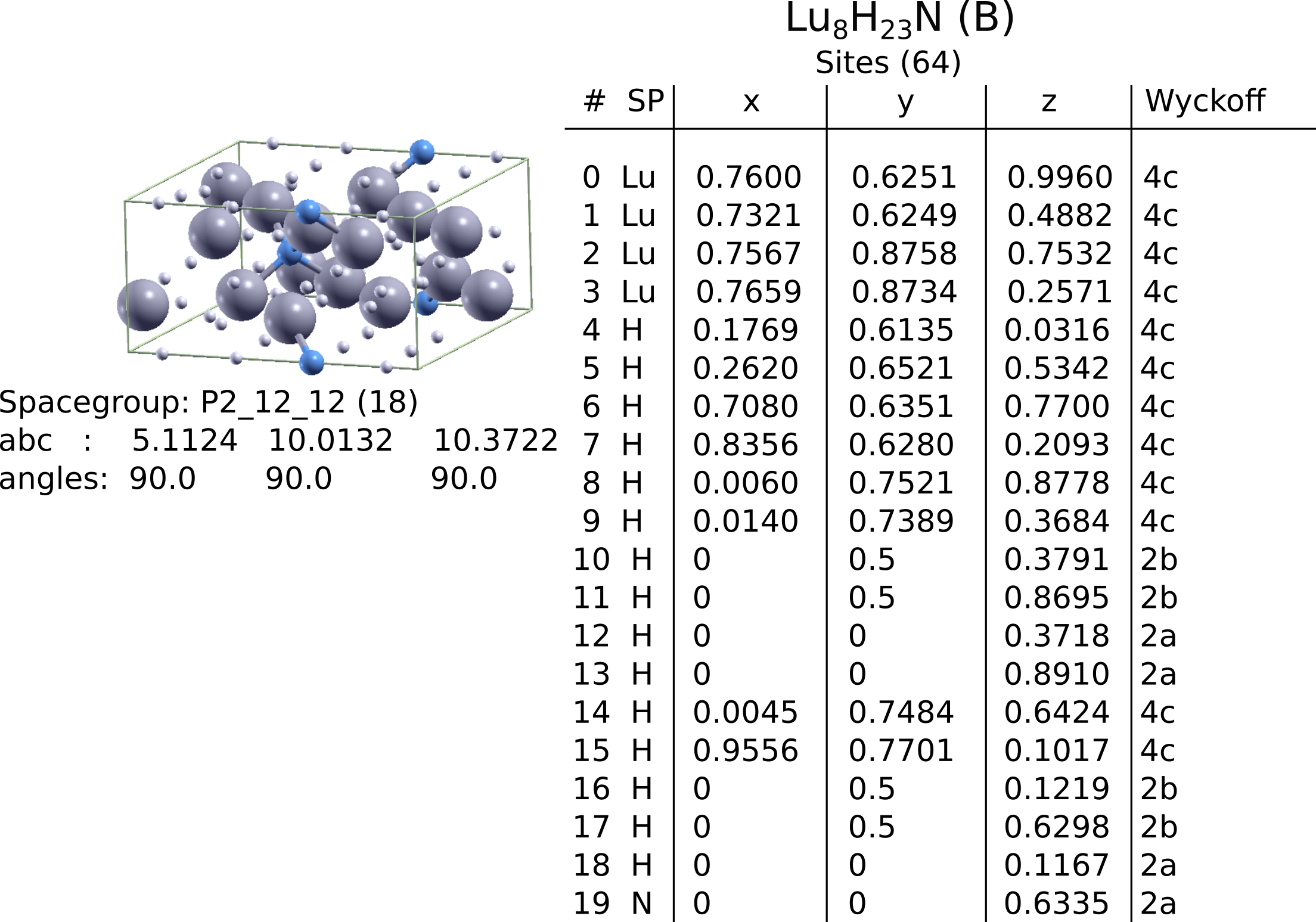}
    \centering \caption{Structure B (Lu$_8$H$_{23}$N), conventional cell parameters, and atomic positions.}
    \label{fig:typeB2_struct}
\end{figure*}

\begin{figure*}
    \centering
    \includegraphics[width=0.85\linewidth]{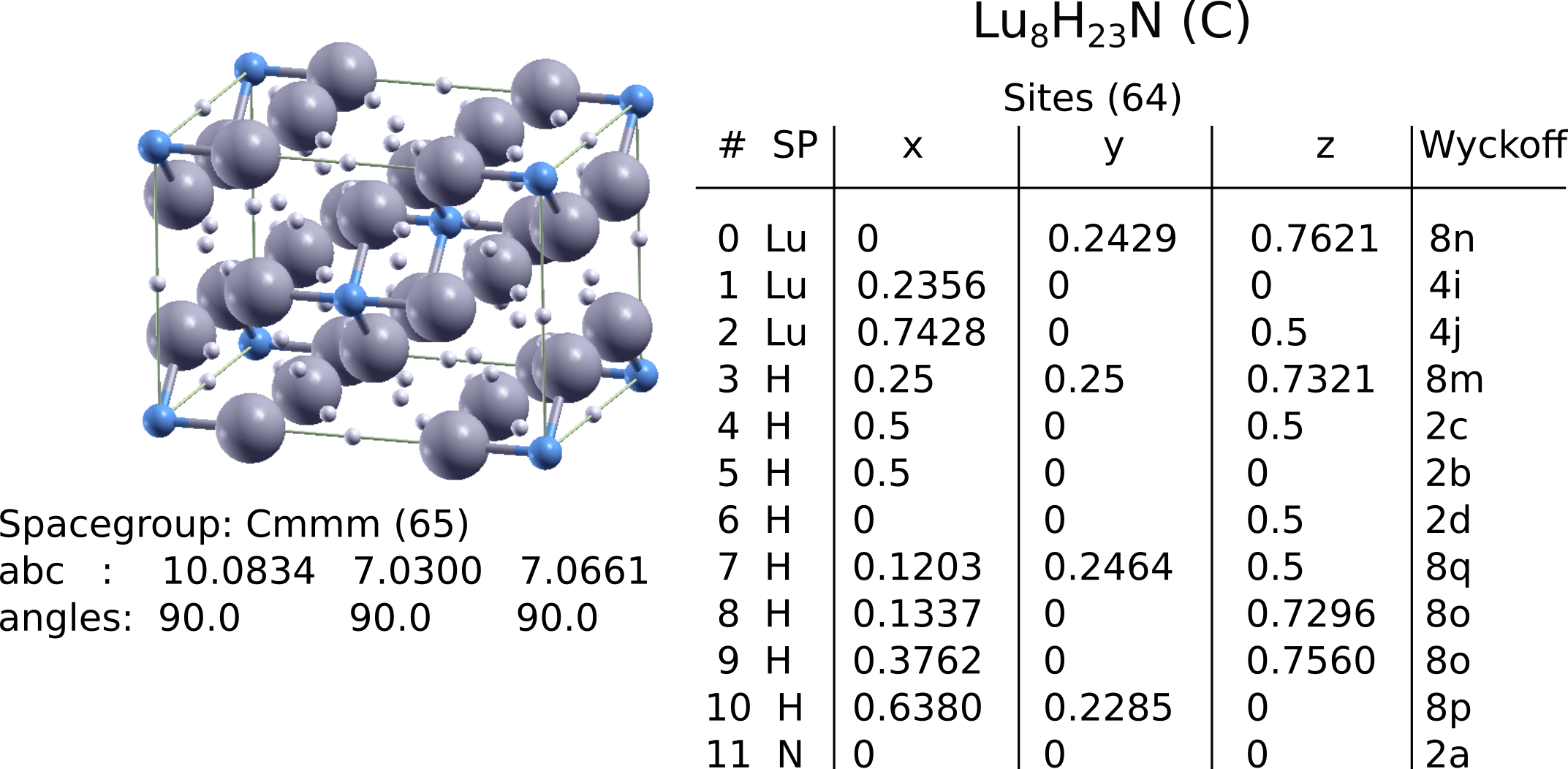}
    \caption{Structure C (Lu$_8$H$_{23}$N), conventional cell parameters, and atomic positions.}
    \label{fig:typeC_struct}
\end{figure*}

\begin{figure*}
    \centering
    \includegraphics[width=0.8\linewidth]{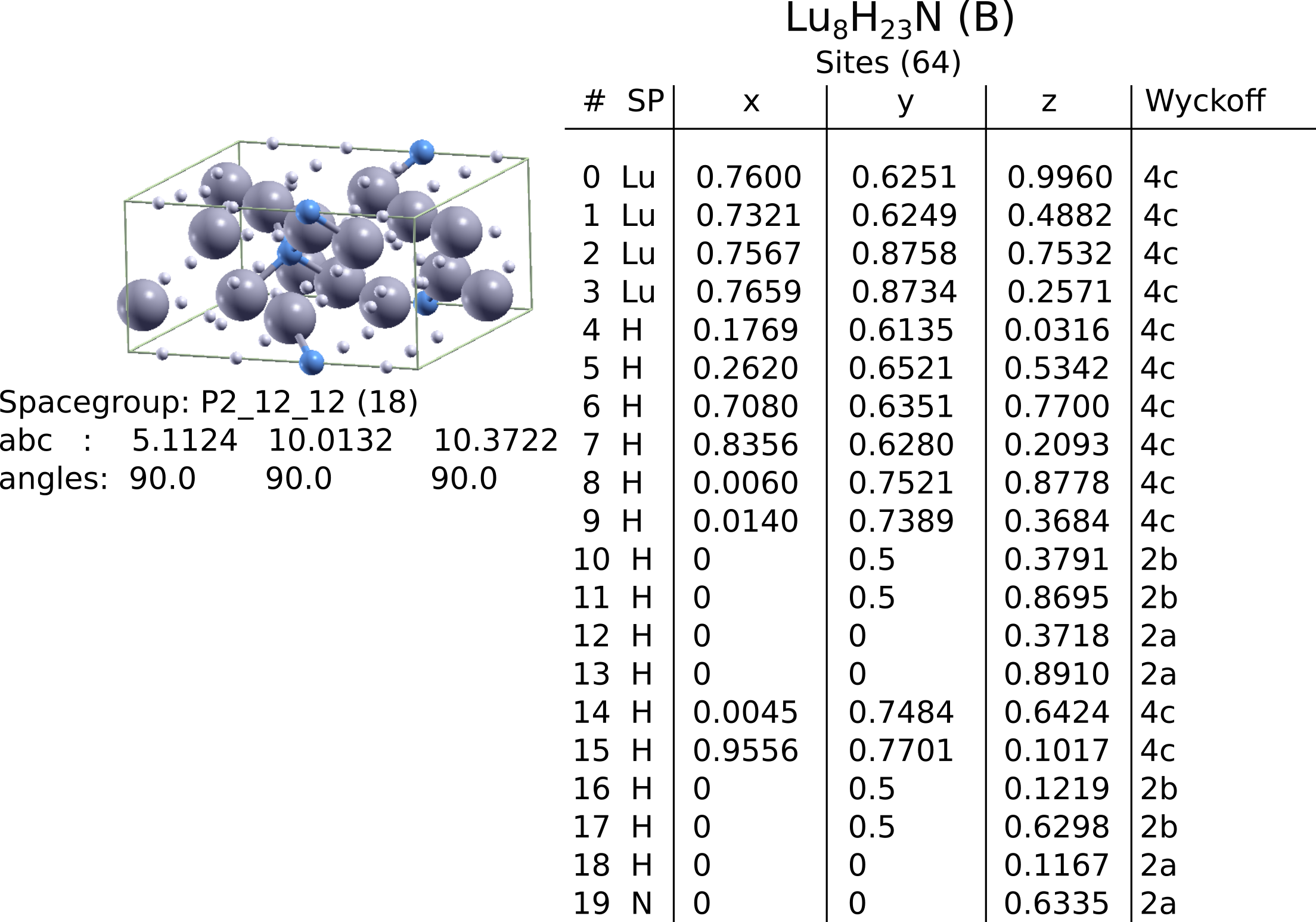}
    \caption{\justifying Structure D (Lu$_8$H$_{23}$N), conventional cell parameters, and atomic positions. The relaxation produced a small deformation away from 90$^o$ angles (89.9998), below the precision used here, so the structure is considered orthorhombic.}
    \label{fig:typeD_struct}
\end{figure*}

\begin{figure*}
    \centering
    \includegraphics[width=0.5\linewidth]{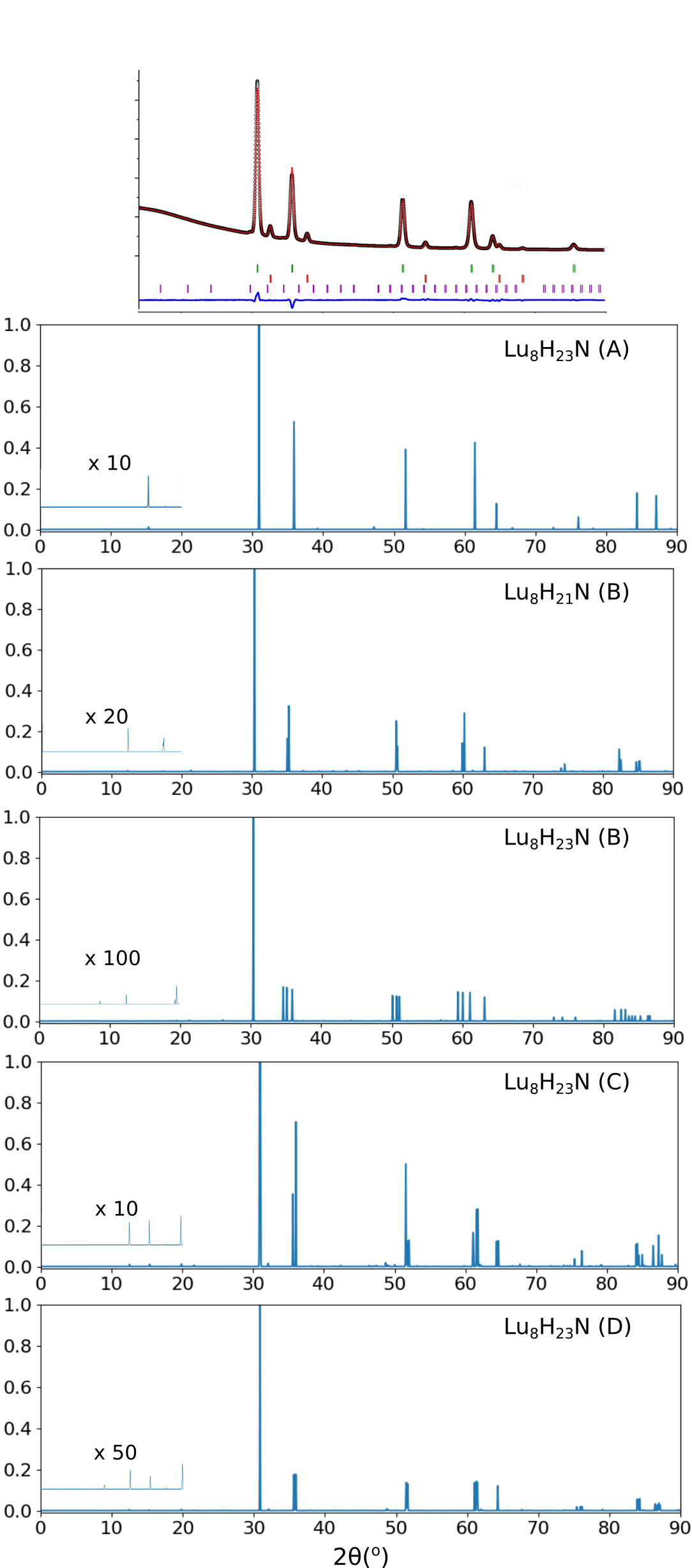}
    \caption{\justifying Calculated x-ray diffraction patterns of predicted structures compared with the experimental results reported in Ref. \cite{dasenbrock2023evidence} (Cu-K$\alpha$ radiation). There is broad agreement between theory and experiment for the principal diffraction peaks (green tick marks in the upper plot for the dominant phase in the experiment). On the other hand, the calculated patterns show the distinct splittings of the principal peaks and low-angle superlattice peaks that were beyond the 2$\theta$ range of the measurements. The red tick marks in the experimental plot correspond to a secondary phase tentatively identified as Lu(H,N) \cite{dasenbrock2023evidence} or LuH \cite{liu2023parent}. The purple tick marks correspond to an additional minor phase (see Ref. \cite{dasenbrock2023evidence}).}
    \label{fig:XRD_compare}
\end{figure*}

\begin{figure*}
    \centering
    \includegraphics[width=1.0\linewidth]{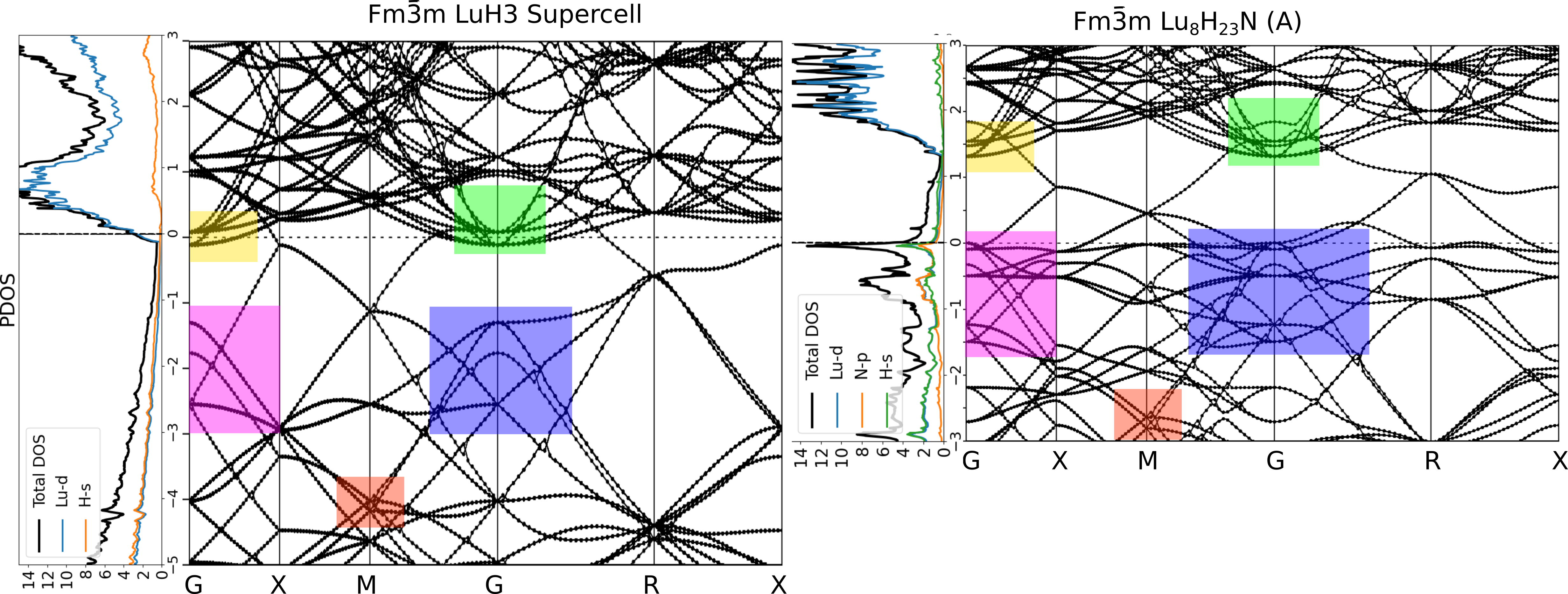}
    \caption{\justifying Band structures of Fm$\overline{3}$m LuH$_3$ (left) and Lu$_8$H$_{23}$N (A) (right) calculated using DFT-PBE. The colored areas highlight various regions that shift fairly uniformly upon introducing N atoms, the incorporation of which gives rise to the additional (flat) bands.}
    \label{fig:LuH3_BS_compare}
\end{figure*}

\begin{figure*}
    \centering
    \includegraphics[width=1.0\linewidth]{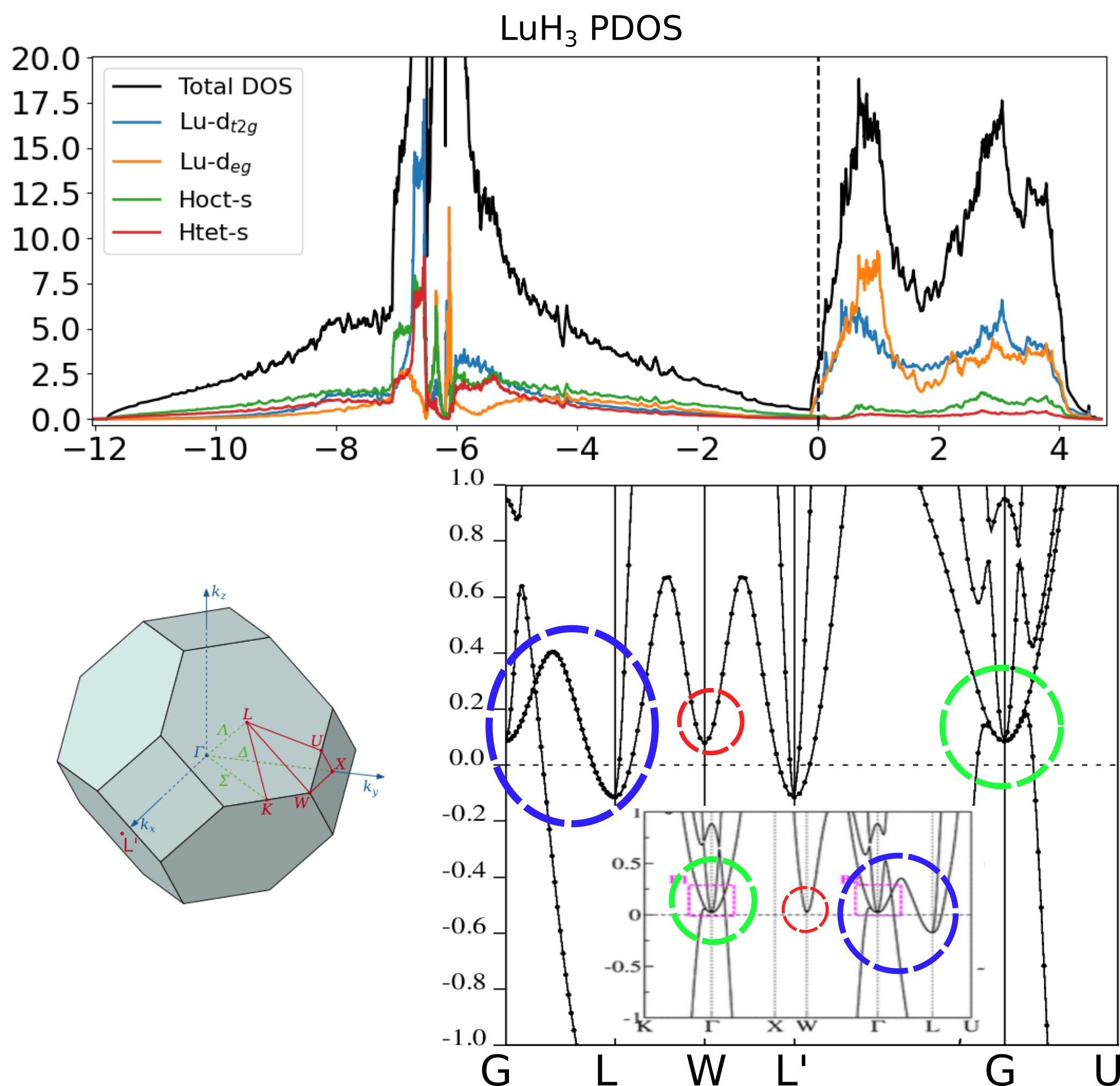}
    \caption{\justifying Top: PDOS of a 128 atom Fm$\overline{3}$m supercell of LuH$_3$ calculated using DFT-PBE. Broadly similar results are presented in Ref. \cite{dasenbrock2023evidence} which also used DFT-PBE but different pseudopotentials. Hydrogen states are mostly below E$_F$, as they tend to strongly hybridize with Lu$_d$ orbitals. Bottom: band structure of Fm$\overline{3}$m LuH$_3$ (primitive cell representation). Similar results were reported in Ref. \cite{sufyan2023topological} using the SCAN functional (inset), where the colored circles indicate comparable regions between the two calculations. Topological features such as flat band regions and Dirac cones are both seen.}
    \label{fig:LuH3_BS_compare_Sufyan}
\end{figure*}

\begin{figure*}
  \centering
    \includegraphics[width=0.9\linewidth]{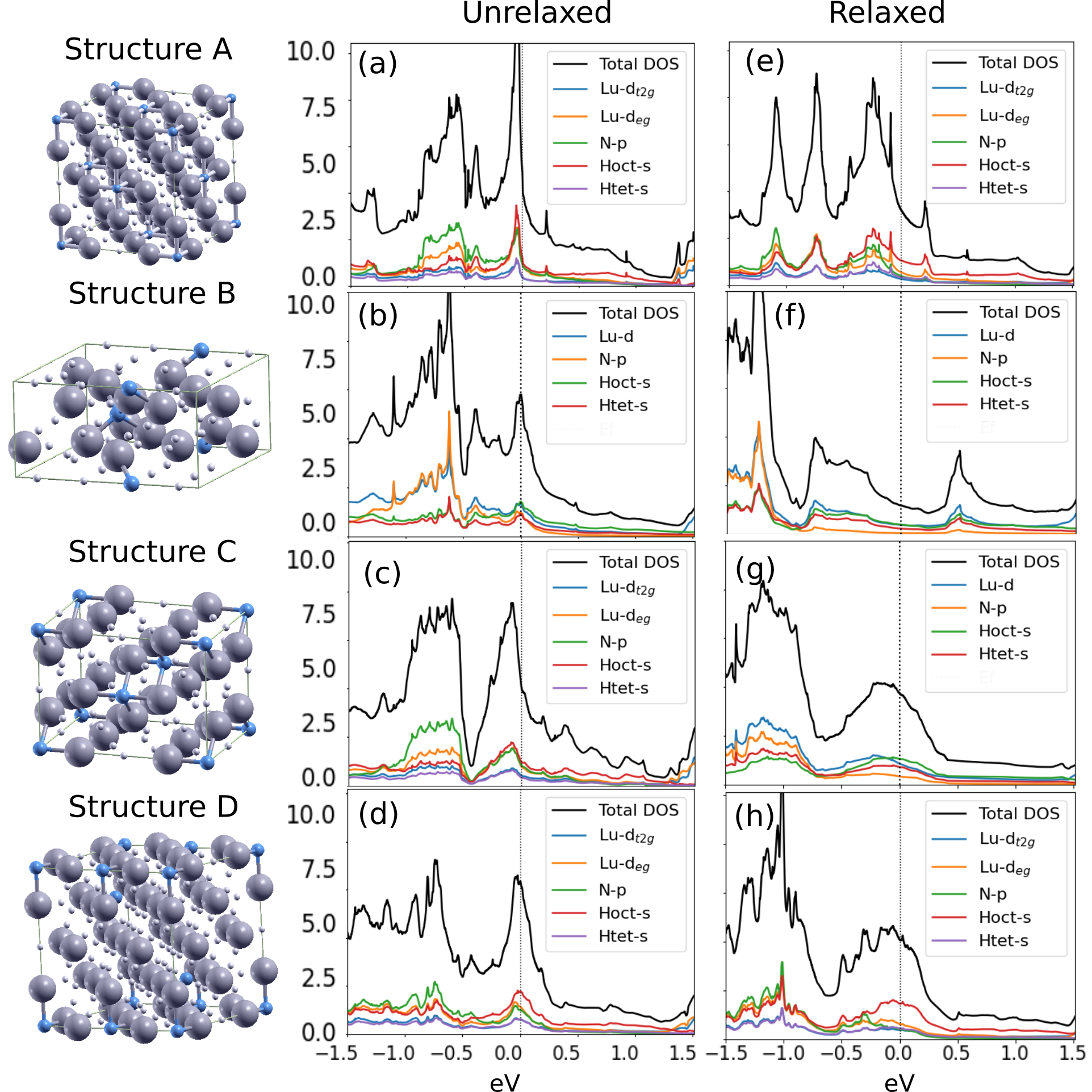}

  \caption{\justifying Comparison of the PDOS of Lu$_8$H$_{23}$N (A-D) using DFT-PBE for unrelaxed (a-d) and relaxed (e-h) structures. Unrelaxed structures are based on  supercells of the parent Fm$\overline{3}$m LuH$_3$ structure with lattice constant $a = $ 4.983 \AA \\ (a) Structure A, with H$^{\text{oct}}_s$, H$^{\text{tet}}_s$ , Lu$_{d_{t2g}}$ and Lu$_{d_{eg}}$ orbitals projected (b) Structure B, which is lower symmetry, the Lu$_{d_{t2g}}$ and Lu$_{d_{eg}}$ projections were not disentangled (c) Structure C (d) Structure D. Structural information of the relaxed unit cells is provided in Figs. \ref{fig:typeA_struct} - \ref{fig:typeD_struct}.}
  \label{fig:8_1_compare}
  
\end{figure*}

\begin{figure*}
    \centering
    \includegraphics[width=1.0\linewidth]{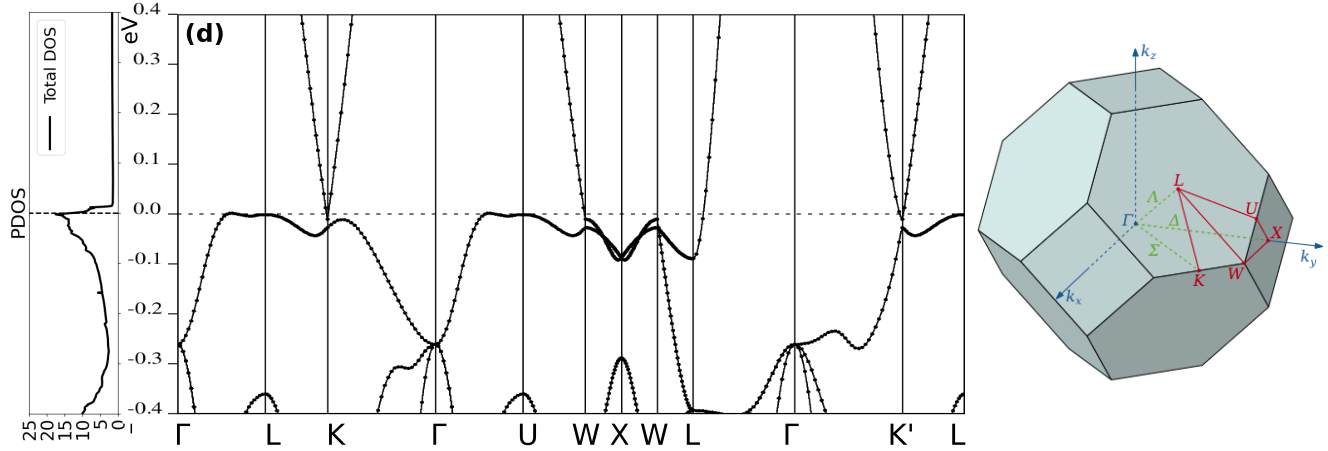}
    \caption{\justifying Detailed DOS and band structure near E$_F$ calculated for the primitive cell representation of Lu$_8$H$_{23}$N (A) using DFT-PBE+U, with U=8.2 eV on Lu$_d$. Dirac cones terminating near E$_F$ at the K and W points are readily observed. }
    \label{fig:NFCC_Prim_BS}
\end{figure*}

\begin{figure*}
    \centering
    \includegraphics[width=1.0\linewidth]{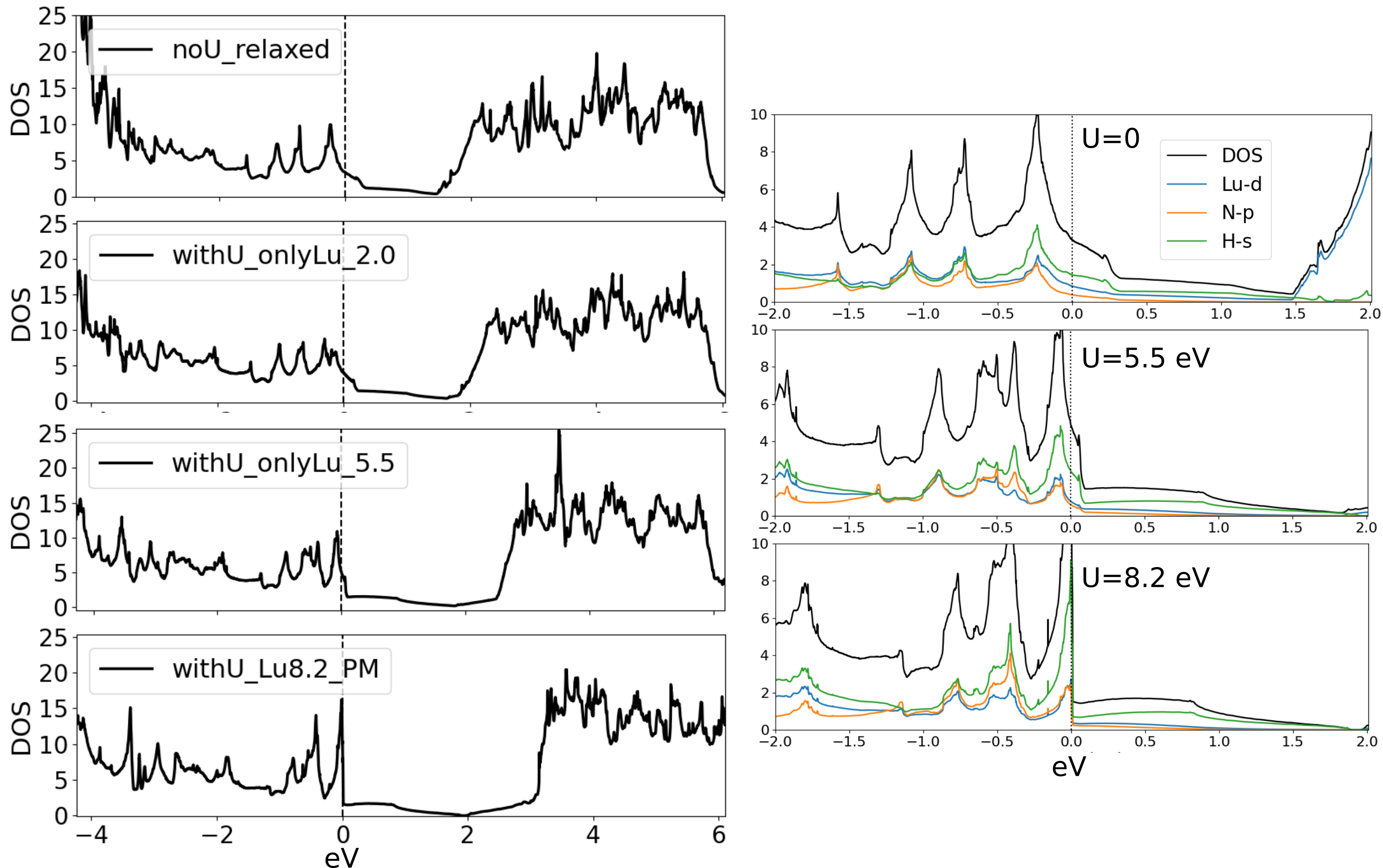}
    \caption{\justifying Effect of U on the DOS (left) and PDOS (right) for Lu$_8$H$_{23}$N (A). U is applied to the Lu$_d$ orbitals. We denote "PM" (paramagnetic) in the bottom DOS because there is also a ferromagnetic solution with a similar DOS (very small splitting), but with energy lowered by less than 1 meV/atom. U = $8.2$ eV is the recommended value in Ref. \cite{topsakal2014accurate} for LuN. The recommended U = 5.5 eV on f-orbitals did not significantly change the states near E$_F$.}
    \label{fig:four_cases}
\end{figure*}

\begin{figure*}
  \centering
    \includegraphics[width=1.0\linewidth]{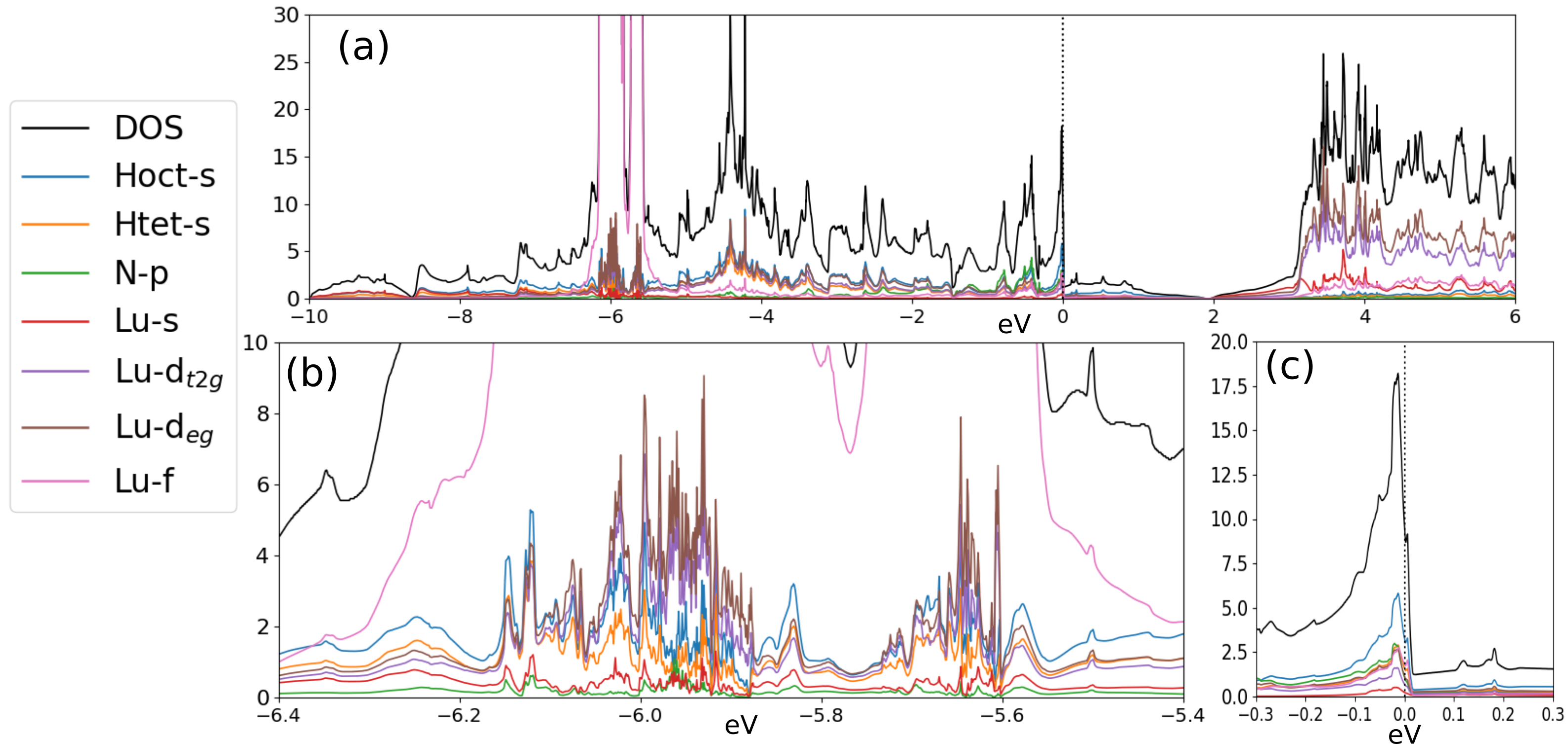}
  \caption{\justifying PDOS of Lu$_8$H$_{23}$N (A). (a) PDOS for an expanded energy range, capturing the increased bandwidth of the Lu$_d$, H$_s$ orbitals relative to Fm$\overline{3}$m LuH$_3$ (see Fig. \ref{fig:LuH3_BS_compare_Sufyan} (b) Detailed PDOS around the Lu$_f$ orbital peak. (c) Detailed PDOS around E$_F$. Both DFT-PBE + U (see main text) and DFT-PBE produce a vHs, though relaxation weakens it and brings it away from E$_F$ using DFT-PBE. The N atoms attract more Lu$_d$ electrons than hydrogen, which drives the energy of occupied hydrogen states closer to E$_F$. Addition of N also has a hole-doping effect. Panel (c) shows that the vHs has primarily N$_p$, Lu$_{d_{eg}}$ and H$^{\text{oct}}_s$ character, with some smaller contributions from the other orbitals. We note here that different projection schemes (QE default atomic projections, vs using PAW projectors) disagree in the weight of the N-p orbitals at E$_F$ for the pure DFT results. Here we use the default QE projections; the PAW projectors yields a higher N$_p$ contribution and lower octahedral H$_s$ contribution at E$_F$. These results do not change when using the vdW-DF3-opt2 functional \cite{chakraborty2020next}.}
  \label{fig:FCCzoom}
\end{figure*}

\begin{figure*}
  \begin{center}
  \includegraphics[width=1.0\linewidth]{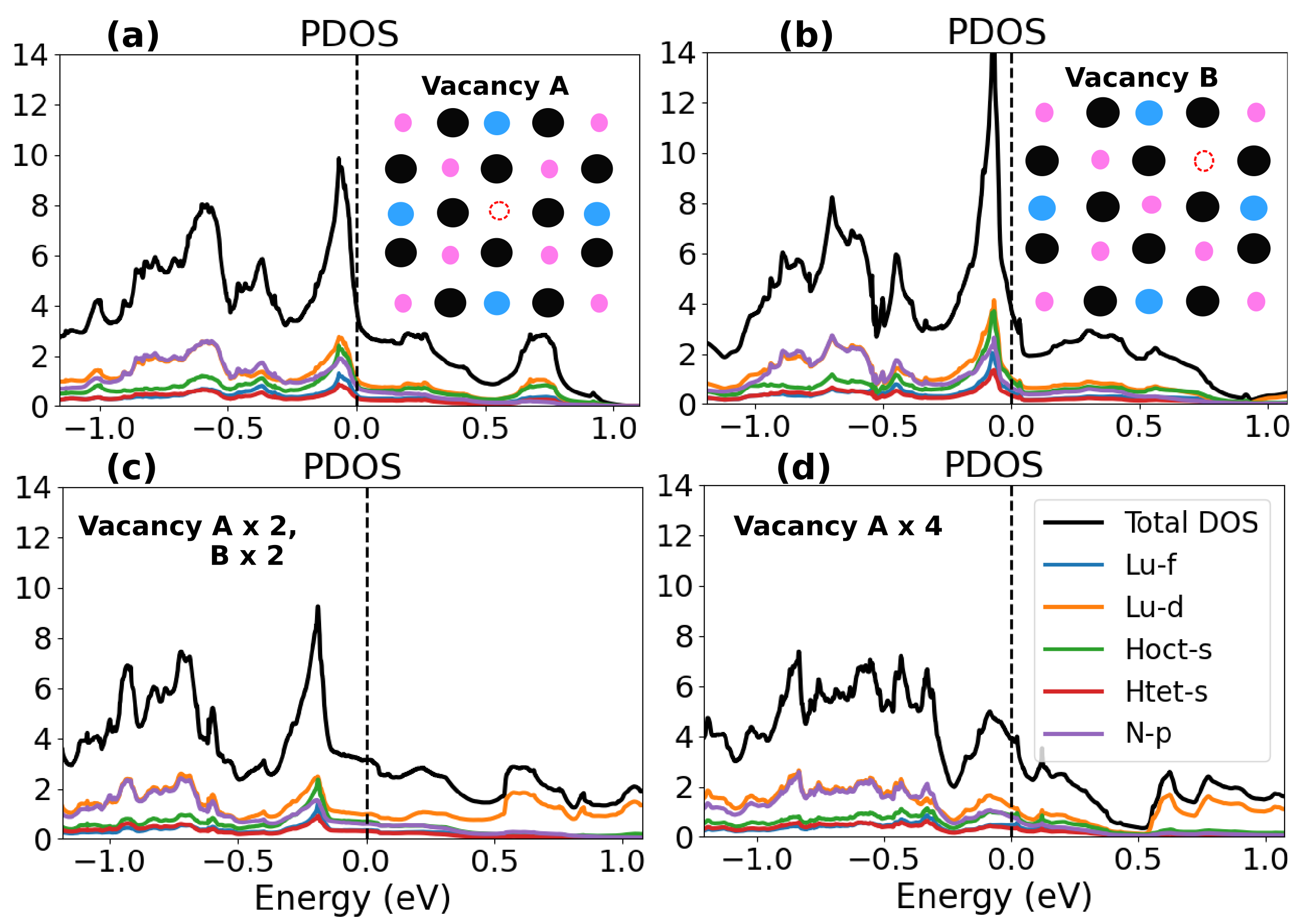}
  \end{center}
  \justifying
  \caption{\justifying (a-d) Effects of different types and numbers of vacancies on the PDOS of  Fm$\overline{3}$m Lu$_8$H$_{23}$N (structure A), unrelaxed (a = 9.975 \AA). With one vacancy the stoichiometry becomes Lu$_{32}$H$_{91}$N. The two types of vacancies considered are shown in the insets of (a,b). While Sun et al. \cite{sun2023effect} showed that replacing H with tetrahedral vacancies in Lu$_8$H$_{23}$N can stabilize the structure; in contrast we study the effects of octahedral vacancies. The location and shape of the vHs is changed based on the ordering and number of vacancies.}
  \label{fig:vaceffect}
\end{figure*}

\begin{figure*}
    \centering
        \includegraphics[width=1.0\linewidth]{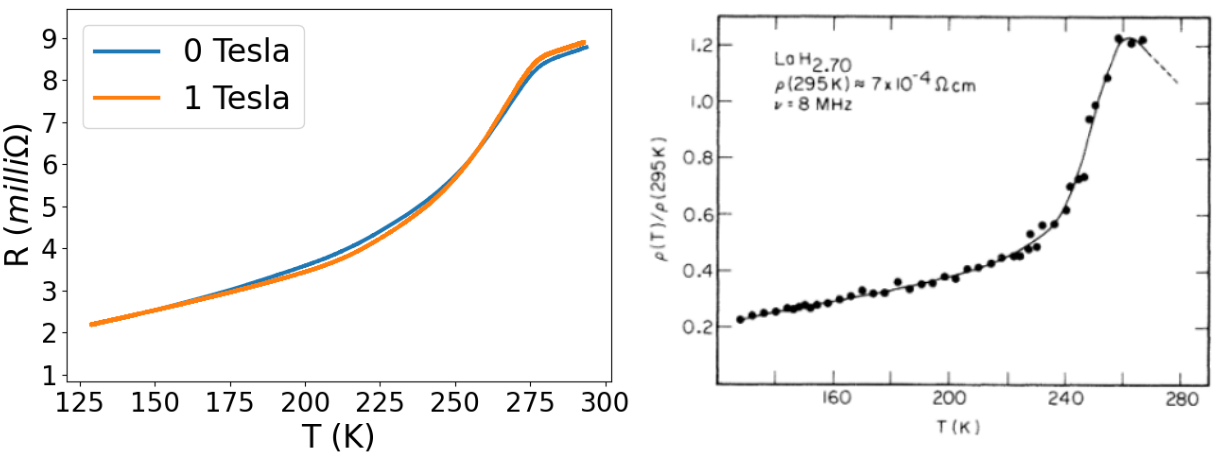}
    \justifying
    \caption{\justifying Uncorrected \emph{R-T} data of Fig.\ S15 for nitrogen-doped lutetium hydride at 1.5 GPa in Ref. \cite{dasenbrock2023evidence} (left), compared to relative \emph{R-T} results for LaH$_{2.70}$ at ambient pressure \cite{shinar1988q} (right).}
    \label{fig:Shinar_vs_Dasenbrock}
\end{figure*}

\begin{figure*}
    \centering
    \includegraphics[width=0.5\linewidth]{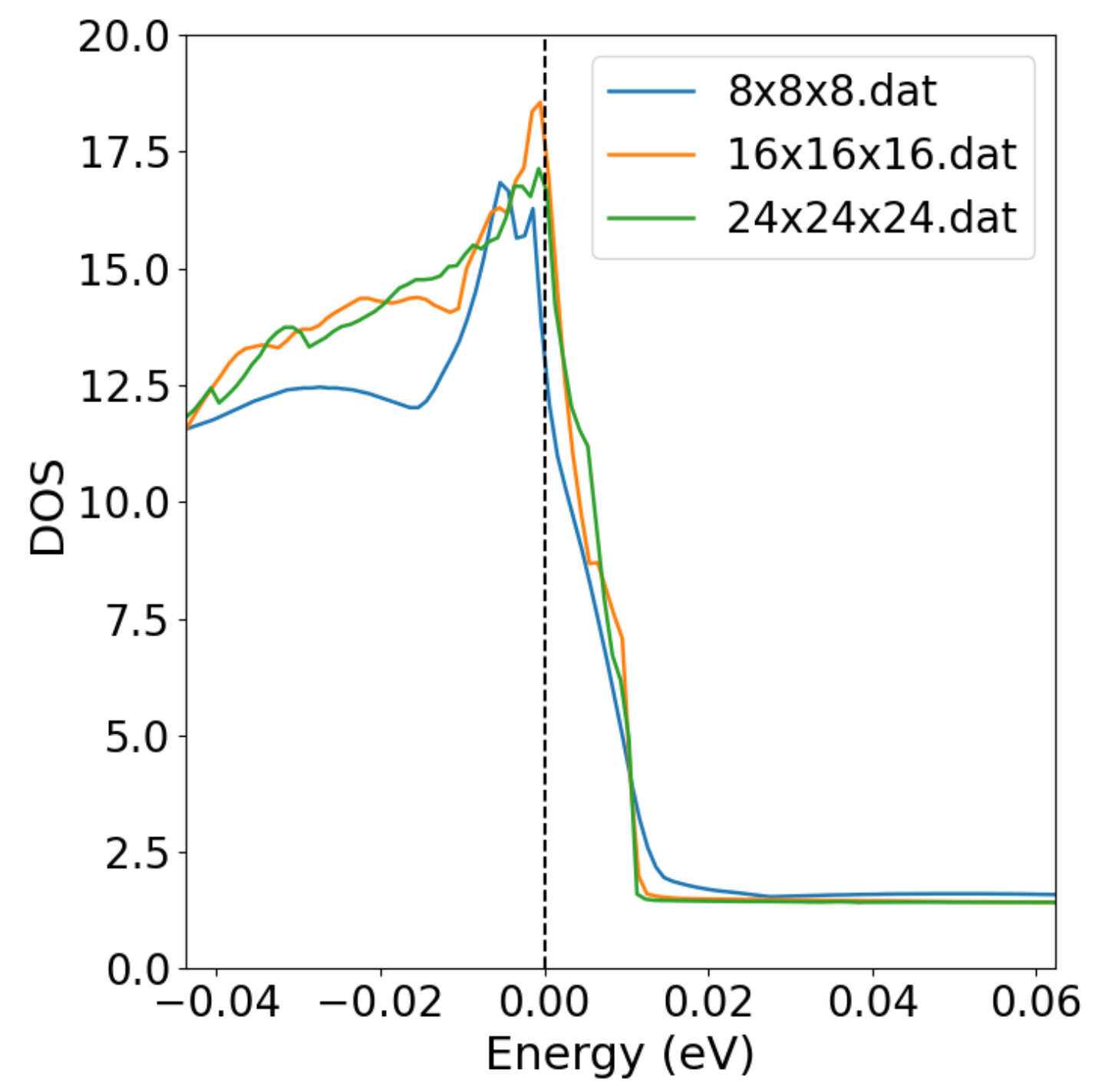}
    \caption{\justifying Mesh dependence of the vHs near E$_F$ calculated for Lu$_8$H$_{23}$N (A) with DFT-PBE+U.}
    \label{fig:kmesh_vHs}
\end{figure*}

\end{document}